\documentclass[conference]{IEEEtran}
\IEEEoverridecommandlockouts
\usepackage{cite}
\usepackage{amsmath,amssymb,amsfonts}
\usepackage{algorithmic}
\usepackage{graphicx}
\usepackage{textcomp}
\usepackage{graphbox}
\usepackage{booktabs}
\usepackage{xcolor}
\def\BibTeX{{\rm B\kern-.05em{\sc i\kern-.025em b}\kern-.08em
    T\kern-.1667em\lower.7ex\hbox{E}\kern-.125emX}}
\begin{document}

\title{Representation of white- and black-box adversarial examples in deep neural networks and humans: A functional magnetic resonance imaging study\\
}

\author{\IEEEauthorblockN{Chihye Han, Wonjun Yoon, Gihyun Kwon, Daeshik Kim}
\IEEEauthorblockA{\textit{Department of Electrical Engineering} \\
\textit{Korea Advanced Institute of Science and Technology}\\
Daejeon, Republic of Korea}
\and
\IEEEauthorblockN{Seungkyu Nam}
\IEEEauthorblockA{\textit{Uiwang R\&D Center} \\
\textit{Hyundai Motor Company}\\
Uiwang, Republic of Korea}
}

\maketitle

\begin{abstract}
The recent success of brain-inspired deep neural networks (DNNs) in solving complex, high-level visual tasks has led to rising expectations for their potential to match the human visual system. However, DNNs exhibit idiosyncrasies that suggest their visual representation and processing might be substantially different from human vision. One limitation of DNNs is that they are vulnerable to adversarial examples, input images on which subtle, carefully designed noises are added to fool a machine classifier. The robustness of the human visual system against adversarial examples is potentially of great importance as it could uncover a key mechanistic feature that machine vision is yet to incorporate. In this study, we compare the visual representations of white- and black-box adversarial examples in DNNs and humans by leveraging functional magnetic resonance imaging (fMRI). We find a small but significant difference in representation patterns for different (i.e. white- versus black-box) types of adversarial examples for both humans and DNNs. However, human performance on categorical judgment is not degraded by noise regardless of the type unlike DNN. These results suggest that adversarial examples may be differentially represented in the human visual system, but unable to affect the perceptual experience.

\end{abstract}

\begin{IEEEkeywords}
fMRI; adversarial example; noise; visual perception; neural network; representational similarity
\end{IEEEkeywords}
\section{Introduction}

State-of-the-art machine vision systems based on deep neural networks (DNNs) achieve remarkable performance in high-level visual tasks such as object recognition \cite{krizhevsky2012imagenet,  simonyan2014very, He_2016_CVPR}. However, existing DNNs are vulnerable to adversarial examples \cite{szegedy2014, goodfellow6572explaining}, which are generated by adding subtle noises that lead a machine classifier, but not a human observer, to misidentify the target image.

While adversarial inputs are a serious security threat for DNNs, they have a negligible influence on humans. Motivated by the difference in degree of robustness, the present work examines the representation of adversarial examples in DNNs and humans.

Specifically, our contributions are as follows:
\begin{itemize}
    \item We obtain feature representations of adversarial examples in the human visual cortex with fMRI and compute their similarity to feature representations produced by hierarchical layers of a DNN using representational similarity analysis \cite{kriegeskorte2008representational}.
    \item Along with white-box examples that exploit access to a DNN structure, adversarial examples with Gaussian black-box noises are presented to humans and a DNN in parallel to investigate their respective neural responses to structured and random noise. 
    \item Concurrently with fMRI measurements, categorization decision performance is recorded in humans to examine whether behavioral judgment and neural representation align for different types of adversaries.
\end{itemize}

The rest of the paper is organized as follows: In Section II, background and prior works are introduced on comparing human and DNN visual systems and adversarial examples. Section III describes methods for fMRI and behavioral experiments. Section IV shows experimental results and analysis. Finally, Section V discusses implications of the results.

\section{Background and Related Work}

\subsection{Comparison of Human and DNN Visual Processing}
DNNs, especially convolutional neural networks (CNNs), have recently achieved human performance in various visual tasks \cite{krizhevsky2012imagenet,  simonyan2014very, He_2016_CVPR}. Like their biologically inspired neural network predecessors \cite{fukushima1982neocognitron, lecun1998gradient,  riesenhuber1999hierarchical}, CNNs share key structural similarities with the ventral visual pathway of the biological brain, including neural receptive fields and hierarchical cortical organization. More importantly, successful CNN variants have shown to exhibit surprising similarities to humans in terms of visual representation and behavior. Neuroimaging studies reported that features from higher layers of DNNs can accurately predict fMRI data from human inferior temporal (IT) cortex and cell recording data from monkey IT, indicating that higher layers of DNNs have obtained similar underlying representations as primate IT for visual object recognition \cite{khaligh2014deep, cadieu2014deep}. Representations from DNNs can be also adapted to reliably model human judgment patterns in letter and image recognition, shape sensitivity, and categorical similarity \cite{testolin2017letter, battleday2017modeling, kubilius2016deep, peterson2017adapting}. Findings that DNNs can closely predict aspects of biological visual processing have suggested their usefulness as a model for biological vision \cite{yamins2016using, doi:10.1146/annurev-vision-082114-035447}.

Despite promising potentials, DNNs exhibit considerable discrepancies from biological vision. Despite initially reported similarities, \cite{Rajalingham240614} showed that the representations of DNNs are significantly non-predictive of primate IT data on an individual image level. Furthermore, DNNs are significantly more susceptible to image distortions such as additive noise, contrast reduction, and reversed brightness \cite{geirhos2017comparing, dodge2016understanding, hosseini2017limitation}. This suggests that DNNs are far less robust than humans, especially in impoverished settings. Finally, psychophysical judgment differences in humans and DNNs \cite{lake2015human, kim2018not} suggest that the underlying visual processing might be substantially different. 

\subsection{Adversarial Examples}
\begin{figure}[tb]
\centerline{\includegraphics[scale=0.4]{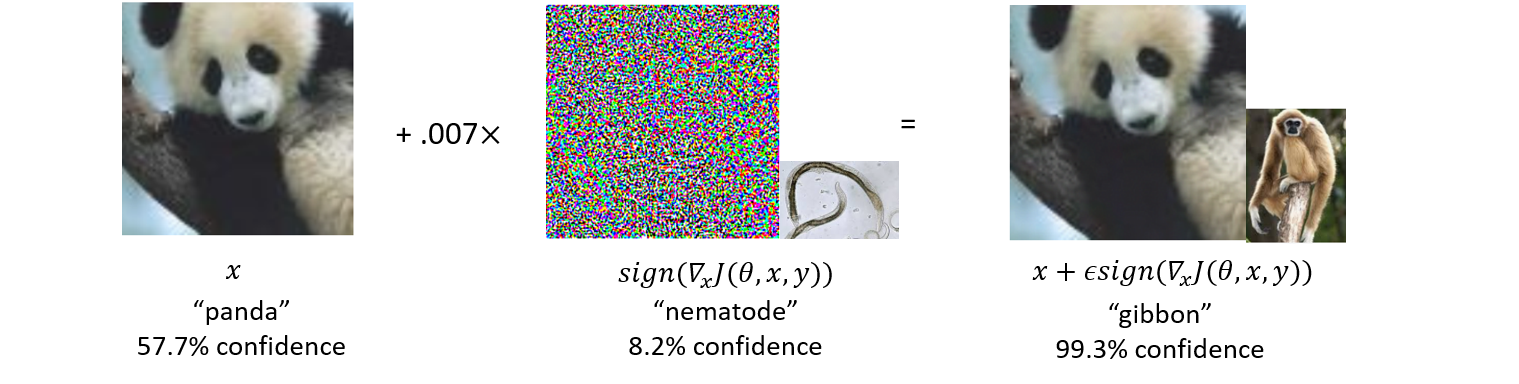}}
\caption{A canonical adversarial example adapted from \cite{goodfellow6572explaining}. An adversarial example is created by generating adversarial noise clipped to be human-imperceptible and adding it to an image of a panda, which is then classified as a gibbon by a deep neural network. See the main text for more detail on adversarial example generation.}
\label{fig1}
\end{figure}

An adversarial example is a case that demonstrates DNNs’ shortcomings to extremes. Adversarial examples are images modified to fool a machine classifier by adding malicious noises to the original \cite{szegedy2014, goodfellow6572explaining}.
A canonical example is shown in Fig. \ref{fig1}, where an image of a panda is misclassified to a gibbon after a human imperceptible noise is added. The generation of an adversarial example can be formally stated as follows, where $x'$ is the adversarial example from the original image $x$ and $\epsilon$ is the noise level:

\begin{equation}
\label{eq:adv}
x' = x \ + \ \epsilon \cdot \eta \text{, where $f(x') \neq y^*$}
\end{equation}

The perturbation noise $\eta$ is constructed by an optimization process that maximizes misclassification of the target image. Such attacks exploit an access to the structure of the neural network they aim to fool and thus are considered white-box attacks (see \textit{Generating Adversarial Stimuli} in the Methods section for details). However, \cite{szegedy2014, PapernotMG16, LiuCLS16} showed that adversarial examples created for one network transfer to other networks with similar structures, enabling black-box attacks. In fact, \cite{LiuCLS16} showed that an adversarial example created based on optimization on multiple networks is more likely to fool another arbitrary network. Arbitrary noises such as Gaussian blur or salt and pepper noises that are inherently independent of any specific network architecture can, by definition, serve as black-box attacks when added to an image until it is misclassified. Adversarial examples also transfer to the real world when captured with cameras or other sensors despite substantial transformations caused by lighting and camera properties \cite{kurakin2016adversarial}. 

While adversarial examples pose a serious security concern for machine classifiers, they are known to have limited impact on humans. \cite{zhang2018} examined the representation of an image of adversarial noise (not of adversarial example) in humans by an fMRI experiment, showing that hierarchical representations of the adversarial noise in humans are increasingly less similar to those of DNN going from low to high layers in the visual cortex. This reaffirms the notion that adversarial noise contains structure meaningful for visual processing of DNNs, but not humans. On the other hand, a psychophysics experiment of \cite{elsayed2018adversarial} suggested that humans, too, are fooled by adversarial examples if exposed to them briefly, i.e. \{71, 63\} ms. It was observed that adversarial examples effective for humans tended to entail visually identifiable modulations in texture, contrast, and edge information, in line with previous accounts that adversarial perturbations sometimes induce semantically meaningful features that are relevant to the target class \cite{athalye2017synthesizing}.

With seemingly equivocal reports of adversarial effects on humans, it is integral to consider the distinction between visual and perceptual representations. Contrary to our subjective impression, our initial sensory representation and final perceptual awareness can well be discrepant. For example, categorical representation in the human IT departs from human judgments such that human categorical judgments, but neither human nor monkey IT representation obtained by fMRI, reflect human-related sub-categorization within the animate class into human vs. nonhuman animals and the inanimate class into natural vs. artificial objects \cite{ 10.3389/fpsyg.2013.00128}. More relevantly, \cite{bracci2017ventral} report that the ventral visual pathway representation measured by fMRI is more prominently guided by animal appearance over animacy, while the reverse is true for human judgment and DNN representation.

In the present work, both visual representations and perceptual performance are considered as we examine fMRI and behavioral patterns of human observers in response to adversarial examples. In addition, effects of white-box and black-box noises are examined symmetrically in humans and their machine counterparts to elucidate whether the neural and behavioral responses to adversarial examples are specific to adversarial noise, as opposed to arbitrary, random noise.

\section{Methods}

\subsection{Stimulus Image}
Stimuli presented to human subjects and a DNN model were adapted from \cite{kiani2007, kriegeskorte2008matching}. The original human fMRI experiment consisted of presenting 96 color images (175 $\times$ 175 pixels) of categorical real-world objects, including animates (faces or bodies of human and nonhuman animals) and inanimates (natural and artificial objects). Time constraint posed by the need to repeat fMRI experiments for several experimental conditions motivated us to exclude the 'human body' and 'nonhuman body' subcategories in our experiment, leaving only 12 images of human face and 12 images of nonhuman face in the animate category. Twelve images from naturalistic and artificial objects in the inanimate category were selected to match per category image count based on non-ambiguity e.g. selecting a single object image over a scenery. The final stimuli consisted of 48 images (see Table \ref{tbl:stim}), resized to 224 $\times$ 224 pixels to enable the use of a DNN pre-trained with images of the same size.

\begin{table}[tb]
\caption{Stimulus Image Set}
\begin{center}
\begin{tabular}[t]{@{}cccc}
\toprule
\textbf{Category} & \textbf{Class} & \textbf{Instance} & \textbf{Examples} \\ \hline
Animate & Human face & 12 & \includegraphics[align=t,height=6mm]{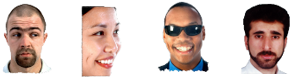} \\ \cline{2-4}  
 & Animal face & 12 &  \includegraphics[align=t,height=6mm]{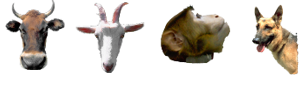} \\ \cline{2-4} 
Inanimate & Natural objects & 12 &  \includegraphics[align=t,height=6mm]{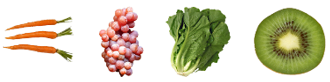} \\ \cline{2-4} 
 & Artificial objects & 12 &  \includegraphics[align=t,height=6mm]{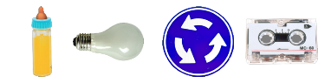}\\         \bottomrule
\end{tabular}
\end{center}
\label{tbl:stim}
\end{table}

\subsection{Generating Adversarial Stimuli}

For our DNN model, we used the PyTorch implementation of the VGG19 network to produce adversarial examples and to compute DNN features \cite{simonyan2014very}. The VGG19 model consists of sixteen convolutional layers and three fully connected layers. The network was pre-trained with 1.2 million labelled images of 1000 categories from ImageNet \cite{imagenet_cvpr09}. 

We generated adversarial images with two types of white-box attacks (with respect to DNN): Projected Gradient Descent (PGD) and Carlini and Wagner (C\&W) attacks. With both white-box attacks, we designated the target class to be the gibbon class in ImageNet ("368: 'gibbon, Hylobates lar'"). For black-box adversarial images, we generated Gaussian noise not designed to deceive a specific target network. With all attacks, we verified that the top one classification result successfully changed to the target class for all 48 stimuli images. A sample image of the generated stimuli in each adversarial condition is provided in Table \ref{tble:example}.

\paragraph{PGD} PGD is a sub-type of a gradient-based attack referred to as fast gradient sign method~(FGSM) \cite{szegedy2014intriguing} in which the adversarial noise is determined by a gradient of loss between the predicted output, $f(x)$, and the ground truth, $y^*$, as follows:

\begin{equation}
\label{eq:fgsm}
x' = x \ + \ \epsilon \cdot \text{sign($\nabla_x$ $J(f(x),y^*)$)}
\end{equation}

PGD is a multi-step variant of FGSM in that it finds the adversarial perturbation by using the same equation as FGSM, but iteratively. The algorithm finds the adversary starting with random $\epsilon$-uniform perturbation clipped in the range of the pixel values of [0, 255]. An image was put into the target boundary by subtracting the sign of a loss between $f(x'_n)$ and $y_{target}$: 
\begin{equation}
\label{eq:pgd}
\begin{array}{cl}
x'_0 = Clip_x(x + uniform(-\epsilon, \epsilon)),  \\
x'_{n+1} = \text{$Clip_{x,\epsilon}$} (x'_{n} \ - \ \alpha \cdot \text{sign($\nabla_x$ $J(f(x'_{n}),y_{target})$)})& \\
\end{array}
\end{equation}
We chose $\alpha$ of 1 and $\epsilon$ of $48$ with $n=20$ steps, but stopped early when the prediction matched the target. 

\paragraph{C\&W} C\&W attack \cite{carliniwagner17} is a strong optimization-based attack in which the adversarial noise is defined with learnable parameters optimized by Adam \cite{Kingma2015Adam}. We used the loss function suggested by the original paper as follows:

\begin{equation}
\label{eq:cw_loss}
f(x) = max(Z(x)_{y^{*}} - \underset{i \neq y^{*}}{max}(Z(x)_i), -\kappa),
\end{equation}
where $Z(x)$ is a logit space through the network given an input $x$ and $\kappa$ is the parameter that controls confidence of finding an adversarial example. We minimized equation (\ref{eq:cw_loss}) using PGD, and we chose $\epsilon$ of $48$ with $100$ iterations and learning rate of $0.004$.

\paragraph{Gaussian} For the black-box attack, we added Gaussian noise with mean and standard deviation of $0.0$ and $0.5$, respectively. We empirically determined the proper value of standard deviation by comparing the resulting level of intensity to other types of noise with the naked eye.

\begin{table}[]
\caption{Example Stimulus Image}
\begin{center}
\begin{tabular}{ccccc}
\hline
\textbf{Condition} & (a) Clean & (b) PGD & (c) C\&W & (d) Gaussian \\
\multicolumn{5}{r}{\includegraphics[align=t,height=15mm]{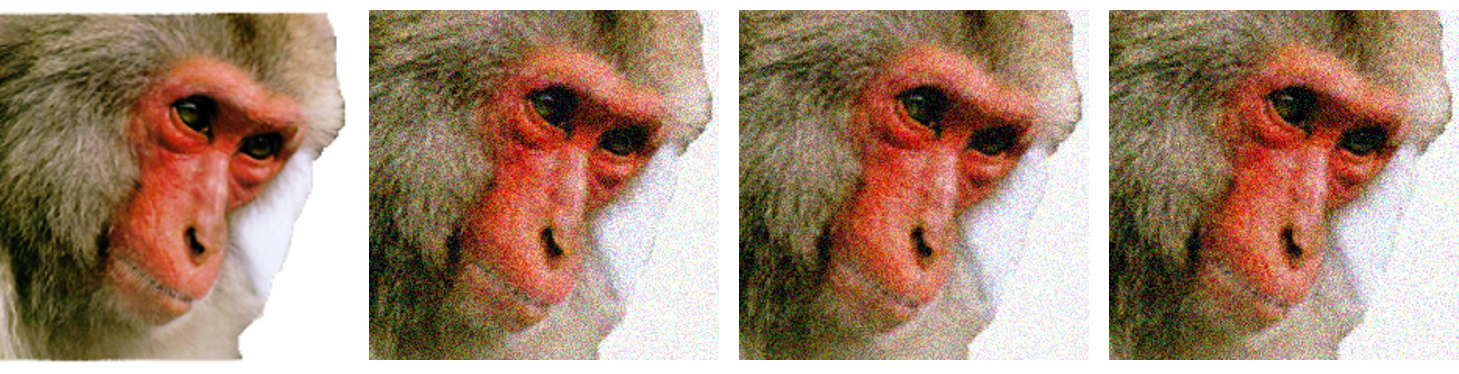}}  \\ \hline
\end{tabular}
\end{center}
\label{tble:example}
\end{table}

\subsection{fMRI Experiment}

\subsubsection{Participants} 
Fourteen healthy subjects were recruited for the study (3 females, mean age 23.68, range 21-30). All subjects had normal or corrected-to-normal visual acuity of 20/40 or above and no neurological or psychiatric history. Subjects provided written informed consent regarding their participation. Experiments were in compliance with the safety guidelines for MRI research and approved by the Institutional Review Board for research involving human subjects at Korea Advanced Institute of Science and Technology.

\subsubsection{MRI Acquisition} 
Experiments were performed with a 12-channel 3T MR scanner (Siemens Magnetom Verio, Germany). The functional images were acquired with a T2*-weighted gradient recalled echo-planar imaging (EPI) sequence (TR, 2,000 ms; TE, 30 ms; flip angle, 90$^\circ$; FOV: 64 $\times$ 64 mm; voxel size, 3 $\times$ 3 $\times$ 3 mm, number of slices, 36). Upon completion of functional imaging, T1-weighted magnetization-prepared rapid-acquisition gradient echo (MPRAGE) images were acquired for normalization purposes (TR, 1,800 ms; TE, 2.52 mx; FA, 9$^\circ$; FOV, 256 $\times$ 256 mm; voxel size, 1 $\times$ 1  $\times$ 1 mm).

Subjects were briefed on MR safety and experimental procedures and guided through a practice run of the behavioral task (see below, \textit{Experimental Design and Tasks}) before entering the scanner. They held a button press handle in each hand for the behavioral task throughout the experiment. Experimental stimuli were presented with MR-compatible video goggles (Nordic Neuro Lab, Norway).

\begin{figure}[t]
\centerline{\includegraphics[scale=0.5]{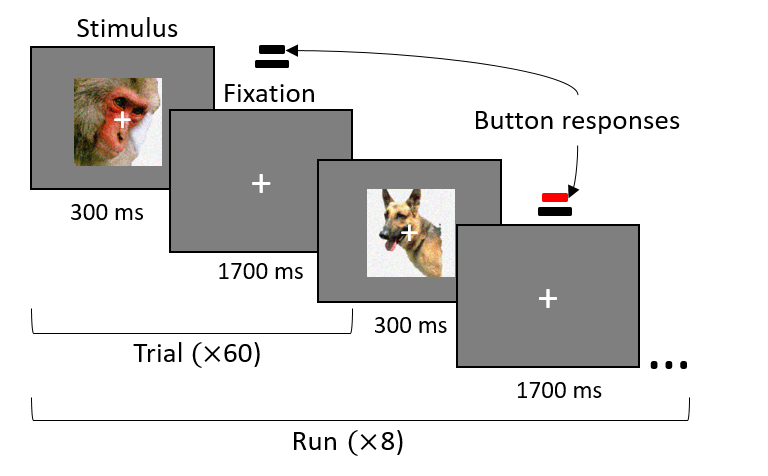}}
\caption{Experimental design. Each subject completed 8 runs, each of which consists of 60 trials (48 visual stimuli in one of 4 adversary conditions and 12 blank trials). Subjects were instructed to perform categorical one-back, pressing the button if subsequent visual stimuli belonged to the same semantic category.}
\label{fig:exp}
\end{figure}

\subsubsection{Experimental Design and Tasks}
The experimental task was programmed with PsychoPy v2.0 for Windows \cite{peirce2007psychopy}. Stimuli images were presented foveally for a duration of 300 ms. The stimulus onset asynchrony (SOA) was 2 s. Null trials were randomly inserted in each run, producing the the effect of jittered interstimulus intervals The resulting SOA for image stimuli trials ranged from 2 s to 12 s. A centered fixation cross appeared throughout the runs.

Each subject completed a total of eight runs, with each run lasting for 4 m and 12 s, all presented on the same day (total duration, 33 m 6 s). Each run belonged to one of four conditions: (1) Clean (unattacked), (2) PGD, (3) C\&W, or (4) Gaussian noise attacked images. Each condition constituted a separate run such that a single run consisted of images from the same condition. Each condition was presented twice in a pseudo-randomly assigned order. Each of the runs randomly presented one of 48 images (visual angle, 9$^\circ$) exactly once, along with randomly interspersed 12 blank null trials showing gray background only. Each run contained 6 s of pre- and post-rest (3 volumes each). Subjects were encouraged to take a break between runs.

Subjects were instructed to fixate on a fixation cross throughout the experiment and to perform a behavioral task of categorical one-back. In this behavioral task, subjects were to press the button with the thumb of their preferred hand if the presented image stimulus belonged to the same category as the previous one (Fig. \ref{fig:exp}). The category used as the basis of the decision was animate (human and animal faces) versus inanimate (naturalistic or artificial objects). Thus, subjects pressed the button if two immediately subsequent stimuli belonged to the animate (or inanimate) class. The button responses were recorded for response accuracy, sensitivity, specificity, and latency, which were respectively calculated as follows:

\begin{equation}
accuracy = \frac{TP+TN}{P+N}
\label{eq:acc}
\end{equation}

\begin{equation}
sensitivity = \frac{TP}{TP+FN}
\label{eq:sen}
\end{equation}

\begin{equation}
specificity = \frac{TN}{FP+TN}
\label{eq:spe}
\end{equation}

\begin{equation}
latency = t_{response_{TP}} - t_{onset_{TP}}
\label{eq:lat}
\end{equation}
$TP$, $TN$, $FP$, and $FN$ indicate the number of true positive, true negative, false positive, and false negative, respectively. $P$ is the total positive case of $TP+FN$, and $N$ is the total negative case of $TN +FP$. t\textsubscript{response\textsubscript{TP}} and t\textsubscript{onset\textsubscript{TP}} refer to the time of response and the time of stimulus onset for a true positive instance, respectively. One-way analysis of variances (ANOVAs) was performed to detect significant effects of the noise type in each of these measures.

\subsubsection{Data Preprocessing}
fMRI data preprocessing was performed using Statistical Parameter Mapping (SPM12, Wellcome Trust Centre for Neuroimaging, London, UK). The first three volumes of each run were discarded automatically during the scanning process for magnetic field stabilization. We performed a rigid body transform motion correction across runs in each subject using the middle volume as a reference. Functional images were directly normalized to the Montreal Neurological Institute (MNI) template (East Asian brains). The normalized images were rewritten at 3mm isometric voxels. No spatial smoothing was applied as recommended for representational similarity analysis. 

\begin{table}[t]
\caption{Regions of Interest (ROI) Definition for Human and DNN}
\begin{center}
\resizebox{\columnwidth}{!}{
\begin{tabular}{ccccc}
\toprule
\textbf{} & \textbf{} & \textbf{Anatomical region} & \textbf{ROI} & \textbf{Abbr.} \\ \hline
\textbf{Human} & 1 & Gyrus fusiformis & \{FG1 FG2 FG3 FG4\} & FG \\
 & 2 & hOC1 & \{hOC1\} & hOC1 \\
 & 3 & hOC2 & \{hOC2\} & hOC2 \\
 & 4 & Ventral extrastriate cortex & \{hOC3d hOC3v\} & hOC3d/4d \\
 & 5 & Dorsal extrastriate cortex & \{hOC4d hOC4v\} & hOC3v/4v \\
 & 6 & Lateral occpital cortex & \{hOC4la hOC4lp\} & hOC4l \\\hline
\textbf{DNN} & 1 &  & \{conv1\_1 conv1\_2\} & conv1 \\
 & 2 &  & \{conv2\_1 conv2\_2\} & conv2 \\
 & 3 &  & \{conv3\_1 conv3\_2 conv3\_3 conv3\_4\} & conv3 \\
 & 4 &  & \{conv4\_1 conv4\_2 conv4\_3 conv4\_4\} & conv4 \\
 & 5 &  & \{conv5\_1 conv5\_2 conv5\_3 conv5\_4\} & con5 \\
 & 6 &  & \{fc1\} & fc1 \\
 & 7 &  & \{fc2\} & fc2 \\
 & 8 &  & \{fc3\} & fc3 \\ \bottomrule
\end{tabular}}
\label{tbl:roi}
\end{center}
\end{table}

\subsubsection{Regions of Interest (ROI) Definition} Beta maps were extracted from normalized functional volumes for regions of interest (ROI). ROIs were generated based on anatomical probability maps provided by SPM Anatomy Toolbox \cite{eickhoff2005new}. A total of 12 maps including V1-4 (i.e. hOC1-4) and fusiform gyrus (i.e. FG) were chosen to represent the visual area (See Table~\ref{tbl:roi}). The number of voxels per mask ranged from 316 to 3331. MarsBaR \cite{brett2002region} was used to produce masks from the probability maps and to extract the masked beta maps. For beta map extraction, we masked functional images obtained 6 s (3 volumes) after the stimuli onset to account for hemodynamic delays. Each of these beta vectors was taken as the neural representation for an image stimulus in each visual area. Extracted beta vectors $b \textsubscript{r, c}$ from ROI $r$ and condition $c$ were further normalized with mean $\mu\textsubscript{r, c}$ and standard deviation $\sigma\textsubscript{r, c}$ before using them as input for the representational similarity analysis:

\begin{equation}
b'_{r,c} = \frac{b_{r,c} - \mu_{r,c}}{\sigma_{r,c}}
\label{norm}
\end{equation}

\subsection{Representational Similarity Analysis}
Representational similarity analysis \cite{kriegeskorte2008representational} is a framework that enables comparisons of representations from different modalities, e.g. computational models and fMRI patterns, by comparing the dissimilarity patterns of the representations. Representations from different modalities are compared by first constructing representational dissimilarity matrices (RDMs). RDM is a square symmetric matrix that contains pairwise (dis)similarity values between response patterns of all stimuli pairs. Using the representational similarity analysis toolbox \cite{nili2014toolbox}, we constructed RDM for every ROI (shown in Table \ref{tbl:roi}) $\times$ condition \{Clean, PGD, C\&W, Gaussian\} $\times$ subject \{Human 1, 2, ... , 14, DNN\}. Given 48 stimuli images used in our experiment, each RDM was a 48 $\times$ 48 matrix containing dissimilarity values between the response patterns (fMRI or DNN features) elicited by two stimuli. The dissimilarity measure was 1 minus the Pearson correlation. 

We then constructed the second-level correlation matrix of RDMs by computing the pairwise similarities of individual RDMs, which visually demonstrates the relatedness of brain and DNN representation patterns from each ROI and condition. The similarity measure was the Kendall's rank correlation coefficient $\tau_A$. 

Finally, we performed statistical inference with a one-sided signed-rank test to assess the degree of relatedness between RDMs. This procedure can be used, for example, to test whether a given computational model ('candidate RDM') explains some brain representation ('reference RDM') better than others, cf. \cite{khaligh2014deep}. In our experiment, we set the subject-average human RDMs in response to the clean stimuli as the reference RDM and related them to other brain RDMs or to DNN RDMs. We repeated the process in reverse, with the ROI average of DNN RDMs in response to clean stimuli as the reference RDM.

\section{Results}

\begin{figure}[t]
\centerline{\includegraphics[scale=0.56]{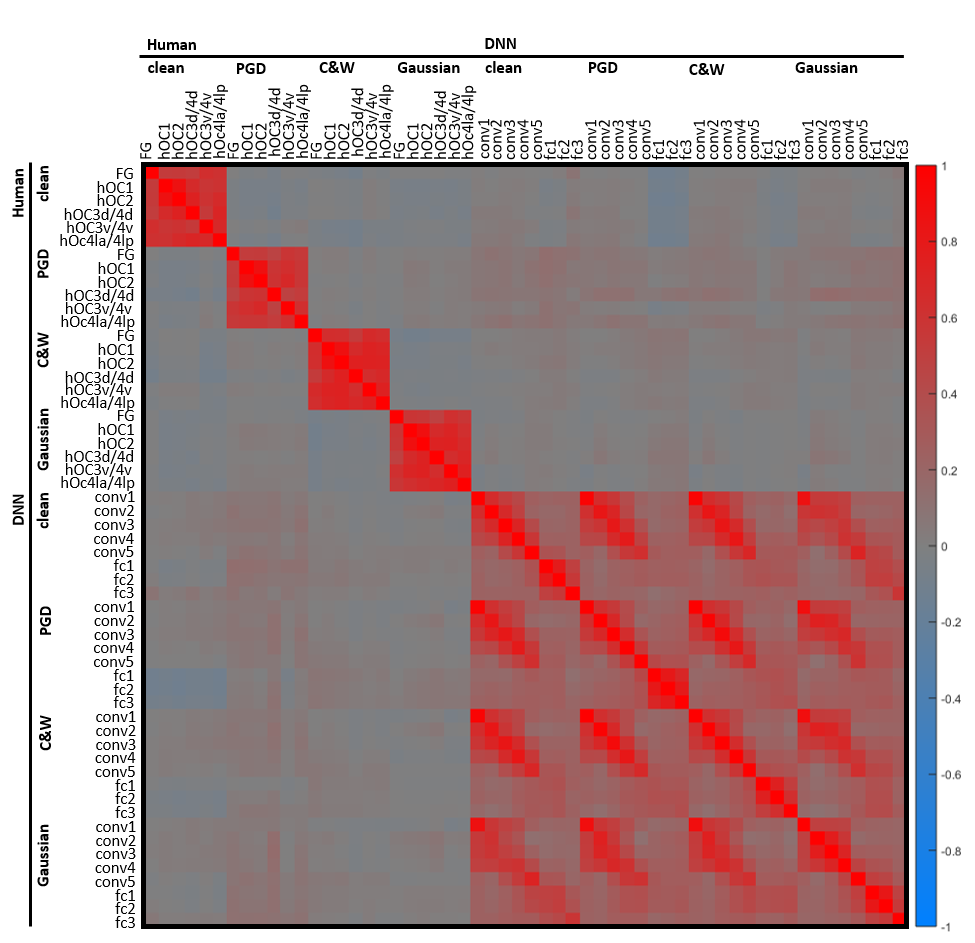}}
\caption{The correlation matrix of subject-average human and DNN RDMs. Each cell represents a pairwise similarity between two RDMs computed with Kendall's rank correlation coefficient $\tau_A$ in the range [-1, 1].}
\label{fig:rdm}
\end{figure}

\subsection{Comparison of Visual Representations}

Fig. \ref{fig:rdm} shows the correlation matrix of subject-average human and DNN RDMs produced from the representational similarity analysis. Each cell represents a Kendall's $\tau_A$ rank correlation coefficient between two RDMs, with each RDM containing response patterns for 48 stimuli (not shown). Each row or column represents correlations between RDMs from a single ROI and the other RDMs. Correlations for the same stimuli condition are grouped together, forming square regions for high within-condition correlations. Correlations between the identical RDMs fill the diagonal with the value of 1.

Visual inspection of the correlation matrix reveals that human RDMs have relatively small within-group correlations (upper left quadrant) compared to DNN RDMs (lower right quadrant), reflecting the difference in noise levels (not corrected).

Within-group correlations in DNNs exhibit distinctive adversary effects reflected by condition-dependent representations in the fully connected layers (fc1-3): Features from fully connected layers of DNNs in response to clean stimuli show a moderate similarity to features of Gaussian-attacked stimuli, but not with those of PGD- or C\&W-attacked stimuli.

Correlations between human and DNN RDMs (lower left quadrant) also show varying patterns by noise type: Human RDMs of clean stimuli (1\textsuperscript{st} column) show small positive correlations with DNN RDMs from convolutional layers (conv1-5), but negative correlations with the higher fc1-3 layers, especially for PGD and C\&W adversarial conditions; Human RDMs of PGD adversary (2\textsuperscript{nd} column) show more positive correlations with DNN RDMs compared to other human conditions; Human RDMs of C\&W and Gaussian adversary (3\textsuperscript{rd} and 4\textsuperscript{th} columns) also show moderate positive correlations with DNN, less in convolutional layers than in fully connected layers.

\begin{figure}[t!]
\centerline{\includegraphics[scale=0.5]{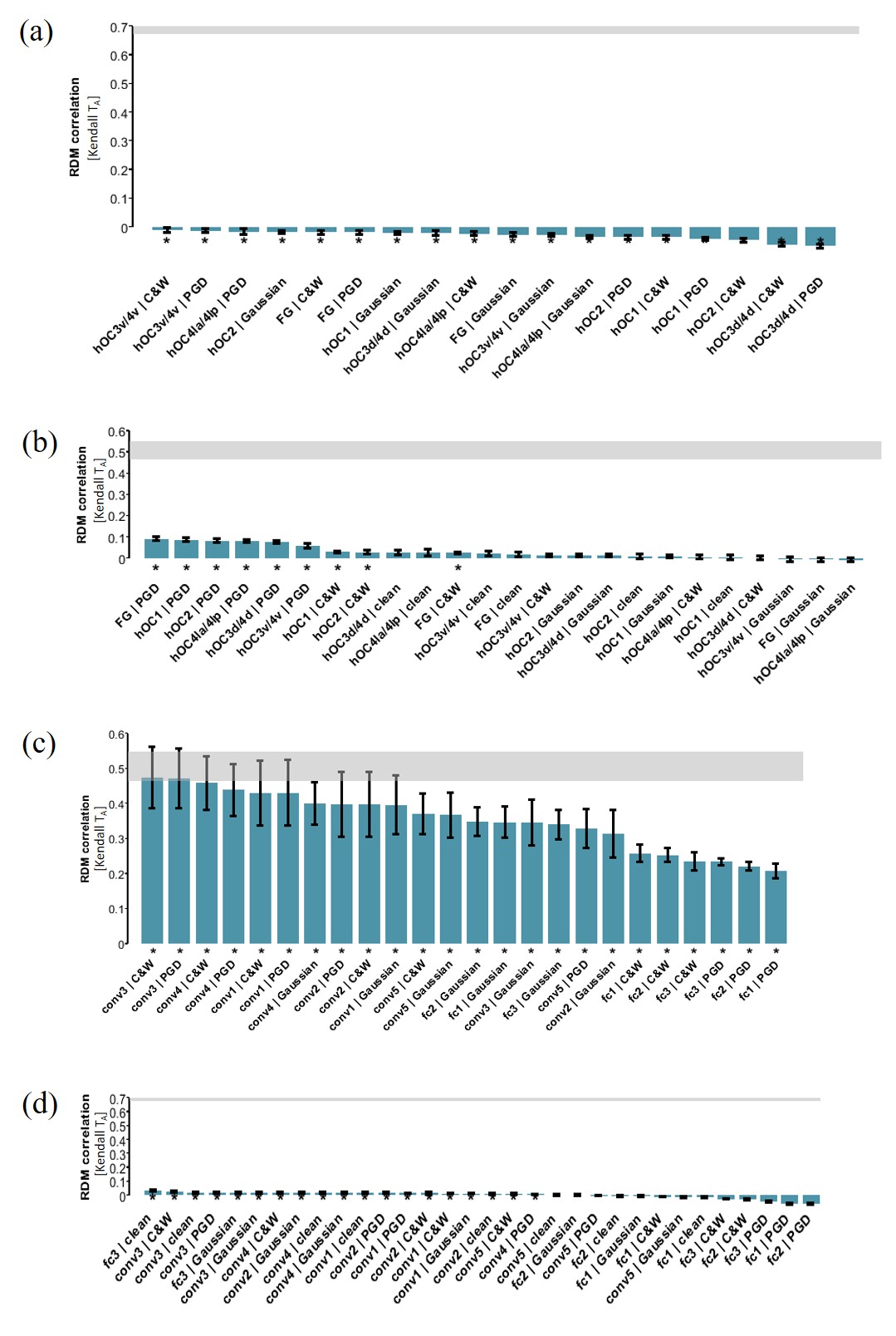}}
\caption{Kendall-$\tau_A$ correlations between (a) human RDMs of adversarial images and human RDMs of clean images (reference), (b) human RDMs and DNN RDMs of clean images (reference), (c) DNN RDMs of adversarial images and DNN RDMs of clean images (reference), (d) DNN RDMs and human RDMs of clean images (reference).}
\label{fig:all}
\end{figure}

The correlation matrix is further supplemented with statistical inference results, shown in Fig. \ref{fig:all}. Here, stars represent significant correlations (p$<$0.05), and the gray box represents noise ceilings.

Fig. \ref{fig:all}(a) shows that the ROI- and subject-average human RDM from the clean condition has small negative correlations with human RDMs from other conditions, all correlations significant. From this, it seems that the human representations of the clean images are significantly different from all the noise conditions, but whether it is so to a different degree by noise type is hard to determine due to the small magnitude of correlations among human RDMs caused by low signal-to-noise ratio of fMRI.

A between-group comparison in Fig. \ref{fig:all}(b) shows that there is a condition-dependent difference in humans by relating each human RDM to the reference DNN RDM in the clean condition, averaged across 8 ROIs. The significance test shows that human RDMs from the white-box adversary conditions (PGD in FG, hOC1, hOC2, hOC4la/4lp, hOC3d/4d, hOC3v/4v and C\&W in hOC1, hOC2, FG), but neither clean nor Gaussian conditions, have significant positive correlations with the reference DNN RDMs from the clean condition.

For comparison to the DNN representations, Fig. \ref{fig:all}(c) shows the relatedness of all DNN RDMs with the same average DNN RDMs from the clean condition as above, where all correlations are significant. Here, representations from conv3/4/1/2 of C\&W and PGD were most similar to those of the reference, followed by conv4/5, fc1-3, fc3 of Gaussian noise. Fully connected layers of PGD and C\&W were more dissimilar to the reference than any others.

Lastly, in Fig. \ref{fig:all}(d), the reference was the average of human RDMs in the clean condition, showing the relatedness of each DNN RDM to the human clean condition reference. The fc3 RDM of the clean condition shows a significant overlap with the reference, followed by conv3 of other conditions as well as fc3 of Gaussian noise. As noted from the correlation matrix, all fully connected layers of the white-box adversary conditions had negative correlations with the human clean reference.

\begin{table}[t!]
\caption{Categorical Judgement Performance in Accuracy (acc), Specificity (spe), and Sensitivity (Sen)}
\begin{center}
\resizebox{\columnwidth}{!}{\begin{tabular}{ccccccccccccc}
\hline
\textbf{} & \textbf{ACC} & \textbf{} & \textbf{} & \textbf{} & \textbf{SPE} & \textbf{} & \textbf{} & \textbf{} & \textbf{SEN} & \textbf{} & \textbf{} & \textbf{} \\ 
\textbf{Subject} & \textbf{Clean} & \textbf{PGD} & \textbf{C\&W} & \textbf{Gaussian} & \textbf{Clean} & \textbf{PGD} & \textbf{C\&W} & \textbf{Gaussian} & \textbf{Clean} & \textbf{PGD} & \textbf{C\&W} & \textbf{Gaussian} \\ \hline
\textbf{1} & 0.989 & 0.968 & 1.000 & 1.000 & 0.957 & 0.981 & 0.958 & 1.000 & 1.000 & 0.941 & 1.000 & 0.978 \\
\textbf{2} & 0.947 & 0.979 & 0.978 & 0.980 & 0.979 & 0.909 & 0.980 & 1.000 & 0.989 & 0.977 & 0.980 & 0.977 \\
\textbf{3} & 0.989 & 0.979 & 0.978 & 1.000 & 1.000 & 0.976 & 1.000 & 1.000 & 0.989 & 1.000 & 1.000 & 0.950 \\
\textbf{4} & 1.000 & 1.000 & 1.000 & 1.000 & 0.989 & 1.000 & 1.000 & 0.968 & 0.979 & 0.976 & 1.000 & 1.000 \\
\textbf{5} & 0.989 & 1.000 & 1.000 & 0.984 & 0.989 & 0.982 & 1.000 & 1.000 & 1.000 & 1.000 & 1.000 & 1.000 \\
\textbf{6} & 0.947 & 0.968 & 0.963 & 0.979 & 0.968 & 0.944 & 0.938 & 1.000 & 0.979 & 0.957 & 0.950 & 1.000 \\
\textbf{7} & 0.989 & 0.989 & 1.000 & 1.000 & 0.989 & 0.975 & 0.976 & 1.000 & 1.000 & 0.981 & 1.000 & 1.000 \\
\textbf{8} & 1.000 & 0.947 & 0.941 & 1.000 & 0.989 & 1.000 & 0.889 & 1.000 & 0.979 & 0.978 & 1.000 & 0.983 \\
\textbf{9} & 0.915 & 0.989 & 0.981 & 1.000 & 0.968 & 0.923 & 0.981 & 1.000 & 0.989 & 0.925 & 0.905 & 1.000 \\
\textbf{10} & 1.000 & 1.000 & 0.833 & 1.000 & 1.000 & 1.000 & 1.000 & 0.848 & 0.840 & 1.000 & 1.000 & 1.000 \\
\textbf{11} & 0.989 & 0.989 & 0.978 & 1.000 & 1.000 & 0.978 & 1.000 & 1.000 & 0.989 & 1.000 & 1.000 & 0.974 \\
\textbf{12} & 0.926 & 0.883 & 0.946 & 0.938 & 0.947 & 0.932 & 0.950 & 0.921 & 0.936 & 0.957 & 0.920 & 0.833 \\
\textbf{13} & 0.968 & 1.000 & 0.942 & 1.000 & 0.989 & 0.967 & 1.000 & 1.000 & 0.968 & 0.981 & 0.971 & 1.000 \\
\textbf{14} & 1.000 & 0.989 & 0.979 & 1.000 & 0.989 & 1.000 & 1.000 & 1.000 & 0.989 & 0.982 & 1.000 & 0.980 \\ \hline
\textbf{Mean} & 0.975 & 0.977 & 0.966 & 0.992 & 0.983 & 0.969 & 0.977 & 0.981 & 0.973 & 0.975 & 0.980 & 0.977 \\
\textbf{Var} & 0.001 & 0.001 & 0.002 & 0.000 & 0.000 & 0.001 & 0.001 & 0.002 & 0.002 & 0.000 & 0.001 & 0.002 \\ \hline
\end{tabular}}
\end{center}
\label{tbl:acc}
\end{table}

\begin{table}[t!]
\caption{Categorical Judgment Performance in Response Latency (In Seconds)}
\begin{center}
\resizebox{0.45\columnwidth}{!}{
    \begin{tabular}{ccccc}
    \hline
    \textbf{Subject} & \textbf{Clean} & \textbf{PGD} & \textbf{C\&W} & \textbf{Gaussian} \\ \hline
    \textbf{1} & 0.766 & 0.805 & 0.721 & 0.790 \\
    \textbf{2} & 0.810 & 0.842 & 0.952 & 0.744 \\
    \textbf{3} & 0.714 & 0.721 & 0.744 & 0.793 \\
    \textbf{4} & 0.477 & 0.514 & 0.514 & 0.493 \\
    \textbf{5} & 0.627 & 0.533 & 0.565 & 0.771 \\
    \textbf{6} & 0.580 & 0.696 & 0.583 & 0.551 \\
    \textbf{7} & 0.511 & 0.539 & 0.509 & 0.527 \\
    \textbf{8} & 0.596 & 0.586 & 0.513 & 0.563 \\
    \textbf{9} & 0.616 & 0.569 & 0.597 & 0.609 \\
    \textbf{10} & 0.598 & 0.683 & 0.643 & 0.672 \\
    \textbf{11} & 0.670 & 0.648 & 0.655 & 0.677 \\
    \textbf{12} & 0.904 & 0.822 & 0.880 & 0.819 \\
    \textbf{13} & 0.683 & 0.770 & 0.747 & 0.810 \\
    \textbf{14} & 0.745 & 0.782 & 0.631 & 0.6 \\ \hline
    \textbf{Mean} & 0.664 & 0.679 & 0.661 & 0.674 \\
    \textbf{Var} & 0.012 & 0.012 & 0.016 & 0.011 \\ \hline
    \end{tabular}
    }
\end{center}
\label{tbl:lat}
\end{table}

\subsection{Behavioral Performance}

Table~\ref{tbl:acc} reports the categorical judgment performance for different conditions in each human subject. The average accuracies were 97.5\%, 97.7\%, 96.6\%, and 99.2\% for Clean, PGD, C\&W, and Gaussian conditions, respectively. There was no significant difference in categorical accuracy among four conditions [F(3, 52)=0.232, p=0.874]. Other performance measures also showed no statistical difference, with average specifities of 98.3\%, 96.9\%, 97.7\%, and 98.1\% [F(3, 52)= 0.331, p=0.803], and average sensitivities of 97.3\%, 97.5\%, 98.0\%, and 97.7\% [F(3, 52)= 0.424, p=0.736].

Table~\ref{tbl:lat} reports the true positive response latency for each  condition. The average response latencies were 0.664 s, 0.679 s, 0.661 s, and 0.674 s for clean, PGD, C\&W, and Gaussian conditions, respectively. No statistical difference was observed [F(3, 52)=0.0689, p=0.976].

\section{Discussion}
Our experimental results found that the presence of adversarial noise, regardless of the type, had no effect on the categorical judgments in human observers. However, in the visual representational space, different types of noise had unique patterns for both human and DNN. In the DNN, white-box adversarial attacks of PGD and C\&W resulted in strongly disrupted patterns in the final, classifying layers of fc1-3, while Gaussian noise had qualitatively different, that is, weaker but more global effects across all layers. The effects of adversarial noise were not as pronounced in human fMRI; However, between-group comparison to DNN features revealed that fMRI data from different noise conditions had distinctive similarity patterns. Notably, neural representations of white-box attacked, but neither clean nor Gaussian noise, had a significant resemblance to the DNN representations. Adversarial-induced neural representations also differed in layer-specific response patterns.

Overall, it was indicated that neural processing in the early visual cortex may represent adversarial noise differently, but humans are somehow unaware of it on the perceptual level, and, as a result, unaffected on the behavioral level. A potential reason for this is that the human visual pathway, but not the machine counterpart, incorporates a correction mechanism located higher in the visual pathway that counters the adversarial effect.

Future work should investigate the role of higher visual areas such as IT in the robust perceptual representation against adversaries. Also, the possibility that gradient-based or other structured noises may be represented differently from random noise by the brain as suggested here should be explored further. Finally, efforts should be made toward building a computational model of the brain that successfully accounts for its representational and behavioral patterns as it could provide a basis for building fundamentally more robust machine vision.

\section*{Acknowledgements}
This work was supported by the Engineering Research Center of Excellence (ERC) Program supported by National Research Foundation (NRF), Korean Ministry of Science \& ICT (MSIT) (Grant No. NRF-2017R1A5A1014708) and by the ICT R\&D program of MSIP/IITP [R-20161130-004520, Research on Adaptive Machine Learning Technology Development for Intelligent Autonomous Digital Companion].

\bibliographystyle{IEEEtran} 
\bibliography{test}

\begin{thebibliography}{10}
\providecommand{\url}[1]{#1}
\csname url@samestyle\endcsname
\providecommand{\newblock}{\relax}
\providecommand{\bibinfo}[2]{#2}
\providecommand{\BIBentrySTDinterwordspacing}{\spaceskip=0pt\relax}
\providecommand{\BIBentryALTinterwordstretchfactor}{4}
\providecommand{\BIBentryALTinterwordspacing}{\spaceskip=\fontdimen2\font plus
\BIBentryALTinterwordstretchfactor\fontdimen3\font minus
  \fontdimen4\font\relax}
\providecommand{\BIBforeignlanguage}[2]{{%
\expandafter\ifx\csname l@#1\endcsname\relax
\typeout{** WARNING: IEEEtran.bst: No hyphenation pattern has been}%
\typeout{** loaded for the language `#1'. Using the pattern for}%
\typeout{** the default language instead.}%
\else
\language=\csname l@#1\endcsname
\fi
#2}}
\providecommand{\BIBdecl}{\relax}
\BIBdecl

\bibitem{krizhevsky2012imagenet}
A.~Krizhevsky, I.~Sutskever, and G.~E. Hinton, ``Imagenet classification with
  deep convolutional neural networks,'' pp. 1097--1105, 2012.

\bibitem{simonyan2014very}
K.~Simonyan and A.~Zisserman, ``Very deep convolutional networks for
  large-scale image recognition,'' \emph{arXiv preprint arXiv:1409.1556}, 2014.

\bibitem{He_2016_CVPR}
K.~He, X.~Zhang, S.~Ren, and J.~Sun, ``Deep residual learning for image
  recognition,'' in \emph{The IEEE Conference on Computer Vision and Pattern
  Recognition (CVPR)}, June 2016.

\bibitem{szegedy2014}
\BIBentryALTinterwordspacing
C.~Szegedy, W.~Zaremba, I.~Sutskever, J.~Bruna, D.~Erhan, I.~Goodfellow, and
  R.~Fergus, ``Intriguing properties of neural networks,'' in
  \emph{International Conference on Learning Representations}, 2014. [Online].
  Available: \url{http://arxiv.org/abs/1312.6199}
\BIBentrySTDinterwordspacing

\bibitem{goodfellow6572explaining}
I.~J. Goodfellow, J.~Shlens, and C.~Szegedy, ``Explaining and harnessing
  adversarial examples (2014),'' \emph{arXiv preprint arXiv:1412.6572}.

\bibitem{kriegeskorte2008representational}
N.~Kriegeskorte, M.~Mur, and P.~A. Bandettini, ``Representational similarity
  analysis-connecting the branches of systems neuroscience,'' \emph{Frontiers
  in systems neuroscience}, vol.~2, p.~4, 2008.

\bibitem{fukushima1982neocognitron}
K.~Fukushima and S.~Miyake, ``Neocognitron: A self-organizing neural network
  model for a mechanism of visual pattern recognition,'' in \emph{Competition
  and cooperation in neural nets}.\hskip 1em plus 0.5em minus 0.4em\relax
  Springer, 1982, pp. 267--285.

\bibitem{lecun1998gradient}
Y.~LeCun, L.~Bottou, Y.~Bengio, and P.~Haffner, ``Gradient-based learning
  applied to document recognition,'' \emph{Proceedings of the IEEE}, vol.~86,
  no.~11, pp. 2278--2324, 1998.

\bibitem{riesenhuber1999hierarchical}
M.~Riesenhuber and T.~Poggio, ``Hierarchical models of object recognition in
  cortex,'' \emph{Nature neuroscience}, vol.~2, no.~11, p. 1019, 1999.

\bibitem{khaligh2014deep}
S.-M. Khaligh-Razavi and N.~Kriegeskorte, ``Deep supervised, but not
  unsupervised, models may explain it cortical representation,'' \emph{PLoS
  computational biology}, vol.~10, no.~11, p. e1003915, 2014.

\bibitem{cadieu2014deep}
C.~F. Cadieu, H.~Hong, D.~L. Yamins, N.~Pinto, D.~Ardila, E.~A. Solomon, N.~J.
  Majaj, and J.~J. DiCarlo, ``Deep neural networks rival the representation of
  primate it cortex for core visual object recognition,'' \emph{PLoS
  computational biology}, vol.~10, no.~12, p. e1003963, 2014.

\bibitem{testolin2017letter}
A.~Testolin, I.~Stoianov, and M.~Zorzi, ``Letter perception emerges from
  unsupervised deep learning and recycling of natural image features,''
  \emph{Nature Human Behaviour}, vol.~1, no.~9, p. 657, 2017.

\bibitem{battleday2017modeling}
R.~M. Battleday, J.~C. Peterson, and T.~L. Griffiths, ``Modeling human
  categorization of natural images using deep feature representations,''
  \emph{arXiv preprint arXiv:1711.04855}, 2017.

\bibitem{kubilius2016deep}
J.~Kubilius, S.~Bracci, and H.~P.~O. de~Beeck, ``Deep neural networks as a
  computational model for human shape sensitivity,'' \emph{PLoS computational
  biology}, vol.~12, no.~4, p. e1004896, 2016.

\bibitem{peterson2017adapting}
J.~C. Peterson, J.~T. Abbott, and T.~L. Griffiths, ``Adapting deep network
  features to capture psychological representations: an abridged report,'' in
  \emph{Proceedings of the 26th International Joint Conference on Artificial
  Intelligence}.\hskip 1em plus 0.5em minus 0.4em\relax AAAI Press, 2017, pp.
  4934--4938.

\bibitem{yamins2016using}
D.~L. Yamins and J.~J. DiCarlo, ``Using goal-driven deep learning models to
  understand sensory cortex,'' \emph{Nature neuroscience}, vol.~19, no.~3, p.
  356, 2016.

\bibitem{doi:10.1146/annurev-vision-082114-035447}
\BIBentryALTinterwordspacing
N.~Kriegeskorte, ``Deep neural networks: A new framework for modeling
  biological vision and brain information processing,'' \emph{Annual Review of
  Vision Science}, vol.~1, no.~1, pp. 417--446, 2015. [Online]. Available:
  \url{https://doi.org/10.1146/annurev-vision-082114-035447}
\BIBentrySTDinterwordspacing

\bibitem{Rajalingham240614}
\BIBentryALTinterwordspacing
R.~Rajalingham, E.~B. Issa, P.~Bashivan, K.~Kar, K.~Schmidt, and J.~J. DiCarlo,
  ``Large-scale, high-resolution comparison of the core visual object
  recognition behavior of humans, monkeys, and state-of-the-art deep artificial
  neural networks,'' \emph{bioRxiv}, 2018. [Online]. Available:
  \url{https://www.biorxiv.org/content/early/2018/01/01/240614}
\BIBentrySTDinterwordspacing

\bibitem{geirhos2017comparing}
R.~Geirhos, D.~H. Janssen, H.~H. Sch{\"u}tt, J.~Rauber, M.~Bethge, and F.~A.
  Wichmann, ``Comparing deep neural networks against humans: object recognition
  when the signal gets weaker,'' \emph{arXiv preprint arXiv:1706.06969}, 2017.

\bibitem{dodge2016understanding}
S.~Dodge and L.~Karam, ``Understanding how image quality affects deep neural
  networks,'' in \emph{Quality of Multimedia Experience (QoMEX), 2016 Eighth
  International Conference on}.\hskip 1em plus 0.5em minus 0.4em\relax IEEE,
  2016, pp. 1--6.

\bibitem{hosseini2017limitation}
H.~Hosseini, B.~Xiao, M.~Jaiswal, and R.~Poovendran, ``On the limitation of
  convolutional neural networks in recognizing negative images,'' in
  \emph{Machine Learning and Applications (ICMLA), 2017 16th IEEE International
  Conference on}.\hskip 1em plus 0.5em minus 0.4em\relax IEEE, 2017, pp.
  352--358.

\bibitem{lake2015human}
B.~M. Lake, R.~Salakhutdinov, and J.~B. Tenenbaum, ``Human-level concept
  learning through probabilistic program induction,'' \emph{Science}, vol. 350,
  no. 6266, pp. 1332--1338, 2015.

\bibitem{kim2018not}
J.~Kim, M.~Ricci, and T.~Serre, ``Not-so-clevr: learning same--different
  relations strains feedforward neural networks,'' \emph{Interface Focus},
  vol.~8, no.~4, p. 20180011, 2018.

\bibitem{PapernotMG16}
\BIBentryALTinterwordspacing
N.~Papernot, P.~D. McDaniel, and I.~J. Goodfellow, ``Transferability in machine
  learning: from phenomena to black-box attacks using adversarial samples,''
  \emph{CoRR}, vol. abs/1605.07277, 2016. [Online]. Available:
  \url{http://arxiv.org/abs/1605.07277}
\BIBentrySTDinterwordspacing

\bibitem{LiuCLS16}
\BIBentryALTinterwordspacing
Y.~Liu, X.~Chen, C.~Liu, and D.~Song, ``Delving into transferable adversarial
  examples and black-box attacks,'' \emph{CoRR}, vol. abs/1611.02770, 2016.
  [Online]. Available: \url{http://arxiv.org/abs/1611.02770}
\BIBentrySTDinterwordspacing

\bibitem{kurakin2016adversarial}
A.~Kurakin, I.~Goodfellow, and S.~Bengio, ``Adversarial examples in the
  physical world,'' \emph{arXiv preprint arXiv:1607.02533}, 2016.

\bibitem{zhang2018}
``Representation of adversarial images in deep neural networks and the human
  brain,'' \emph{Conference on Cognitive Computational Neuroscience 2018.
  Archived at https://ccneuro.org/2018/proceedings/1066.pdf}, 2018.

\bibitem{elsayed2018adversarial}
G.~Elsayed, S.~Shankar, B.~Cheung, N.~Papernot, A.~Kurakin, I.~Goodfellow, and
  J.~Sohl-Dickstein, ``Adversarial examples that fool both computer vision and
  time-limited humans,'' in \emph{Advances in Neural Information Processing
  Systems}, 2018, pp. 3914--3924.

\bibitem{athalye2017synthesizing}
A.~Athalye and I.~Sutskever, ``Synthesizing robust adversarial examples,''
  \emph{arXiv preprint arXiv:1707.07397}, 2017.

\bibitem{10.3389/fpsyg.2013.00128}
\BIBentryALTinterwordspacing
M.~Mur, M.~Meys, J.~Bodurka, R.~Goebel, P.~Bandettini, and N.~Kriegeskorte,
  ``Human object-similarity judgments reflect and transcend the primate-it
  object representation,'' \emph{Frontiers in Psychology}, vol.~4, p. 128,
  2013. [Online]. Available:
  \url{https://www.frontiersin.org/article/10.3389/fpsyg.2013.00128}
\BIBentrySTDinterwordspacing

\bibitem{bracci2017ventral}
S.~Bracci, I.~Kalfas, and H.~O. de~Beeck, ``The ventral visual pathway
  represents animal appearance over animacy, unlike human behavior and deep
  neural networks,'' \emph{bioRxiv}, p. 228932, 2017.

\bibitem{kiani2007}
\BIBentryALTinterwordspacing
R.~Kiani, H.~Esteky, K.~Mirpour, and K.~Tanaka, ``Object category structure in
  response patterns of neuronal population in monkey inferior temporal
  cortex,'' \emph{Journal of Neurophysiology}, vol.~97, no.~6, pp. 4296--4309,
  2007, pMID: 17428910. [Online]. Available:
  \url{https://doi.org/10.1152/jn.00024.2007}
\BIBentrySTDinterwordspacing

\bibitem{kriegeskorte2008matching}
N.~Kriegeskorte, M.~Mur, D.~A. Ruff, R.~Kiani, J.~Bodurka, H.~Esteky,
  K.~Tanaka, and P.~A. Bandettini, ``Matching categorical object
  representations in inferior temporal cortex of man and monkey,''
  \emph{Neuron}, vol.~60, no.~6, pp. 1126--1141, 2008.

\bibitem{imagenet_cvpr09}
J.~Deng, W.~Dong, R.~Socher, L.-J. Li, K.~Li, and L.~Fei-Fei, ``{ImageNet: A
  Large-Scale Hierarchical Image Database},'' in \emph{CVPR09}, 2009.

\bibitem{szegedy2014intriguing}
C.~Szegedy, W.~Zaremba, I.~Sutskever, J.~Bruna, D.~Erhan, I.~Goodfellow, and
  R.~Fergus, ``Intriguing properties of neural networks,'' in
  \emph{International Conference on Learning Representations (ICLR)}, 2014.

\bibitem{carliniwagner17}
N.~Carlini and D.~A. Wagner, ``Towards evaluating the robustness of neural
  networks,'' in \emph{2017 {IEEE} Symposium on Security and Privacy, {SP}
  2017, San Jose, CA, USA, May 22-26, 2017}, 2017.

\bibitem{Kingma2015Adam}
J.~B. Diederik P.~Kingma, ``Adam: A method for stochastic optimization,'' in
  \emph{International Conference on Learning Representations (ICLR)}, 2015.

\bibitem{peirce2007psychopy}
J.~W. Peirce, ``Psychopy—psychophysics software in python,'' \emph{Journal of
  neuroscience methods}, vol. 162, no. 1-2, pp. 8--13, 2007.

\bibitem{eickhoff2005new}
S.~B. Eickhoff, K.~E. Stephan, H.~Mohlberg, C.~Grefkes, G.~R. Fink, K.~Amunts,
  and K.~Zilles, ``A new spm toolbox for combining probabilistic
  cytoarchitectonic maps and functional imaging data,'' \emph{Neuroimage},
  vol.~25, no.~4, pp. 1325--1335, 2005.

\bibitem{brett2002region}
M.~Brett, J.-L. Anton, R.~Valabregue, J.-B. Poline \emph{et~al.}, ``Region of
  interest analysis using an spm toolbox,'' in \emph{8th international
  conference on functional mapping of the human brain}, vol.~16, no.~2.\hskip
  1em plus 0.5em minus 0.4em\relax Sendai, 2002, p. 497.

\bibitem{nili2014toolbox}
H.~Nili, C.~Wingfield, A.~Walther, L.~Su, W.~Marslen-Wilson, and
  N.~Kriegeskorte, ``A toolbox for representational similarity analysis,''
  \emph{PLoS computational biology}, vol.~10, no.~4, p. e1003553, 2014.

\end{thebibliography}


\begin{thebibliography}{00}
    \bibitem{b1} G. Eason, B. Noble, and I. N. Sneddon, ``On certain integrals of Lipschitz-Hankel type involving products of Bessel functions,'' Phil. Trans. Roy. Soc. London, vol. A247, pp. 529--551, April 1955.
    \bibitem{b2} J. Clerk Maxwell, A Treatise on Electricity and Magnetism, 3rd ed., vol. 2. Oxford: Clarendon, 1892, pp.68--73.
    \bibitem{b3} I. S. Jacobs and C. P. Bean, ``Fine particles, thin films and exchange anisotropy,'' in Magnetism, vol. III, G. T. Rado and H. Suhl, Eds. New York: Academic, 1963, pp. 271--350.
    \bibitem{b4} K. Elissa, ``Title of paper if known,'' unpublished.
    \bibitem{b5} R. Nicole, ``Title of paper with only first word capitalized,'' J. Name Stand. Abbrev., in press.
    \bibitem{b6} Y. Yorozu, M. Hirano, K. Oka, and Y. Tagawa, ``Electron spectroscopy studies on magneto-optical media and plastic substrate interface,'' IEEE Transl. J. Magn. Japan, vol. 2, pp. 740--741, August 1987 [Digests 9th Annual Conf. Magnetics Japan, p. 301, 1982].
    \bibitem{b7} M. Young, The Technical Writer's Handbook. Mill Valley, CA: University Science, 1989.
    \end{thebibliography}

\end{document}